\pdfoutput=1 
\documentclass{JINST}
\usepackage{subfigure}          

\newcommand{\degree}{\ensuremath{^\circ}}

\title{From Hybrid to CMOS Pixels ... a possibility for LHC's pixel future?}

\author{N. Wermes\\
\llap{}Physikalisches Institut, University of Bonn,\\
  Nussallee 12, 53115 Bonn, Germany\\
E-mail: \email{wermes@uni-bonn.de}}

\abstract{Hybrid pixel detectors have been invented for the LHC to make tracking and vertexing possible at all in LHC's radiation
intense environment. The LHC pixel detectors have meanwhile very successfully fulfilled their promises and R\&D for 
the planned HL-LHC upgrade is in full swing, targeting even higher ionising doses and non-ionising fluences.
In terms of rate and radiation tolerance hybrid pixels are unrivaled. But they have disadvantages as well, most notably 
material thickness, production complexity, and cost. Meanwhile also active pixel sensors (DEPFET, MAPS) have become real pixel 
detectors but they would by far not stand the rates and radiation faced from HL-LHC. 
New MAPS developments, so-called DMAPS (depleted MAPS) which are full CMOS-pixel structures with charge collection in a depleted region
have come in the R\&D focus for pixels at high rate/radiation levels. This goal can perhaps be realised exploiting HV technologies,
high ohmic substrates and/or SOI based technologies. The paper covers the main ideas and some encouraging results from prototyping R\&D, not hiding the difficulties.
}


\begin{document} 

\section{Introduction}
Pixel detectors as large scale tracking devices have first been developed, built and have been used for the LHC detectors.
They have had a very large impact on the physics performance of the LHC experiments, in the proton-proton collision 
experiments ATLAS \cite{ATLAS-pixel} and CMS \cite{CMS-pixel} operating at very high 
rates and in a very harsh radiation environment, as well as in ALICE \cite{ALICE-pixel}, optimised for heavy ion collisions.
They provide precise tracking and vertexing thus enabling the identification of long-lived heavy particles such as 
$\tau$-leptons and mesons containing heavy quarks ($\tau$-, b-, and c-tagging).    
The conceptual architecture of the LHC pixel detectors is that of so-called hybrid pixels \cite{pixel_book}, in which the pixellated sensor-diode
and the R/O-chip are separate entities.   

While hybrid pixels are so far the only viable concept to cope with the rate and radiation environment in LHC pp-experiments, other concepts combining 
some R/O elements or even full R/O circuitry within the sensing element (pixel diode) have been developed and brought to a state of maturity. Examples are DEPFET pixels
\cite{kemmer_lutz1990} chosen for the pixel detector of the Belle II experiment, to be installed in 2017, and MAPS (Monolithic Active Pixel Sensors 
\cite{Dierickx1998,MAPS-epi_Turchetta:2001}) used and operated since 2014 in the STAR experiment at RHIC \cite{Contin:2015gfa}. MAPS pixels are also planned for the 
ALICE Inner Tracker upgrade \cite{Rossegger:2011wha}. 
Table~\ref{tab:accelerators} compares particle rates and radiation levels at different collider accelerators. It is evident that the demands are highest for LHC and HL-LHC,
larger by several orders of magnitude than at other colliders. So far only hybrid pixel detectors can cope with these levels. At other colliders, however, the 
rate and radiation levels are much reduced such that the mentioned alternative pixel detectors are attractive, offering better spatial resolution and less material.
\begin{table}\label{tab:accelerators}
\begin{center}
\begin{tabular}{|l|c|c|c|c|c|}
\hline
  			& Luminosity		&  BX time 		& particle rate 		& NIEL fluence 				& ion. dose \\
 			& cm$^2$ s$^{-1}$	& ns				 & kHz/mm$^2$ 		& n$_{eq}$/cm$^2$/lifetime 	& Mrad/lifetime \\  \hline
LHC 		&  $10^{34}$	        & 25				 & 1000		 		& 2 $\times $10$^{15}$ 		& 79				 \\  \hline
HL-LHC 		&  $10^{35}$	        & 25				 & 10\,000		 		& 2 $\times $10$^{15}$ 		& >500				 \\  \hline
LHC Heavy Ions	&  6\,$\times 10^{27}$	        & 20 000		 & 10		 	& $> \, $10$^{13}$ 		& 0.7				 \\  \hline
RHIC			&  8\, $\times 10^{27}$	                & 110	 & 3.8		 	&  few \,        10$^{12}$ 		& 0.2				 \\  \hline
SuperKEKB	&  10$^{35}$	                & 2	 & 400		 	&  $\sim \, 3 \times 10^{12}$ 		& 10				 \\  \hline
ILC			&  10$^{34}$	                & 250	 & 350		 	&  $10^{12}$ 		& 0.4			 \\  \hline
\end{tabular}
\end{center}
\caption{Rate and radiation conditions at some particle colliders  in comparison. 
The assumed lifetimes are 7 years, 10 years, 5 years for LHC/ HL-LHC, ILC, and others, respectively. The rates and radiation levels are quoted at respective 
locations of pixel detectors positioned closest to the interaction point.}
\end{table}

In this contribution, the most recent advances in hybrid pixel development are described in the next section. Then, in the third section, new monolithic pixel 
technologies aiming to cope with LHC-like radiation and rate levels are addressed, for which R\&D has started only a few years ago.

\section{Hybrid pixel detectors at LHC pp-Expertiments}
\subsection{The signal and the noise}
\begin{figure}
\centering
\includegraphics[width=0.55\linewidth]{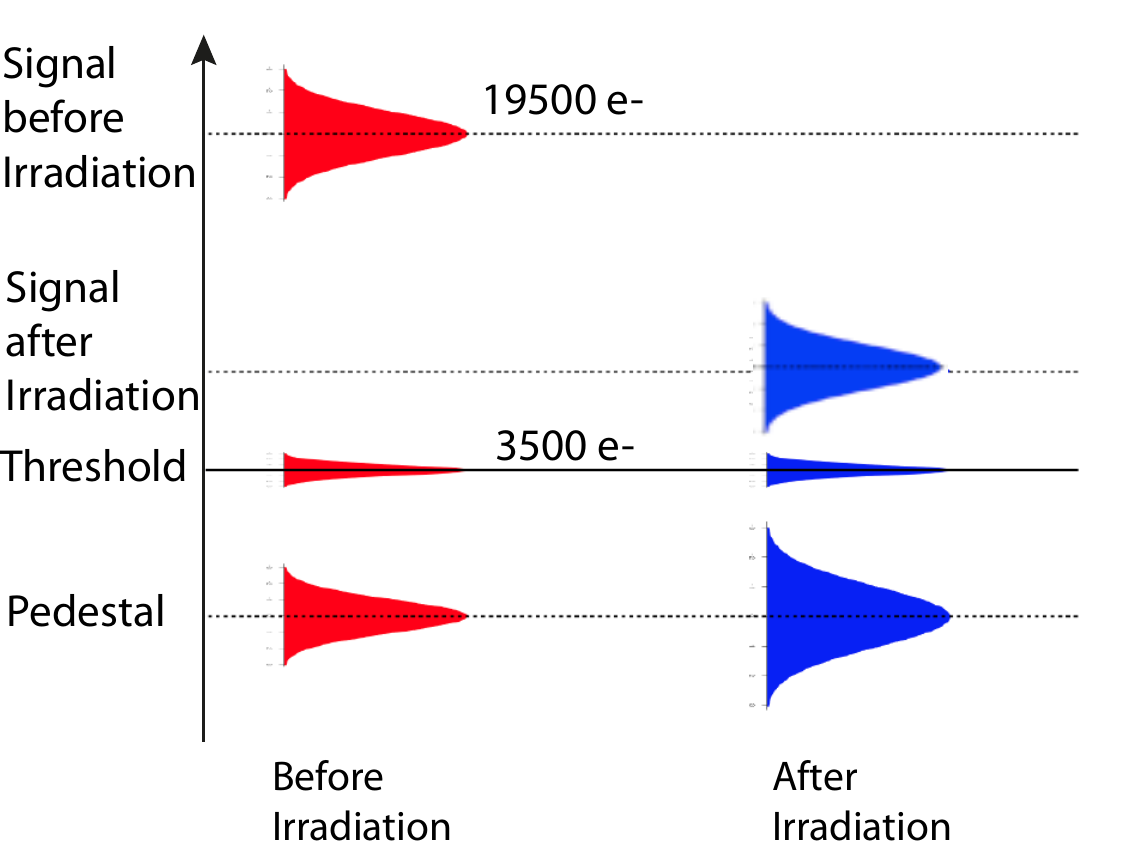}
\caption{Signal and noise situation for a typical pixel detector}
\label{fig:thres_noise} 
\end{figure}

Figure \ref{fig:thres_noise} shows the typical situation faced by pixel detectors operation at LHC in pp-collisions.
In a typical 250\,$\mu$m thick silicon sensor close to 20000 e/h pairs are produced by a minimum ionising particle.
The discriminator threshold is typically set to 3500\,e$^-$. The signal is broadened (dominantly) by the electronics noise, the size
of which can be measured by the width of the pedestal distribution. The probability of a fake hit, however, created when a noise hit
crosses a channel's threshold, is governed by both the
noise and the threshold spread (added in quadrature). After irradiation (up to 2$\times$10$^{15}$ n$_{eq}/$cm$^2$)
the S/N and signal-over-threshold situation becomes much tighter, because the signal is smaller due to bulk damage  
and the noise is larger due to increased leakage current. Fake hits thus become much more likely. In addition the chip 
electronics circuitry may fail due to ionising radiation in the order of 80 Mrad, which causes transistor threshold shifts.  
Note that these conditions become more severe by a factor of roughly 10 at the planned LHC upgrade HL-LHC (see table \ref{tab:accelerators}).

\subsection{IBL - The first pixel upgrade in ATLAS}
\begin{figure}
\centering
\includegraphics[width=0.99\linewidth]{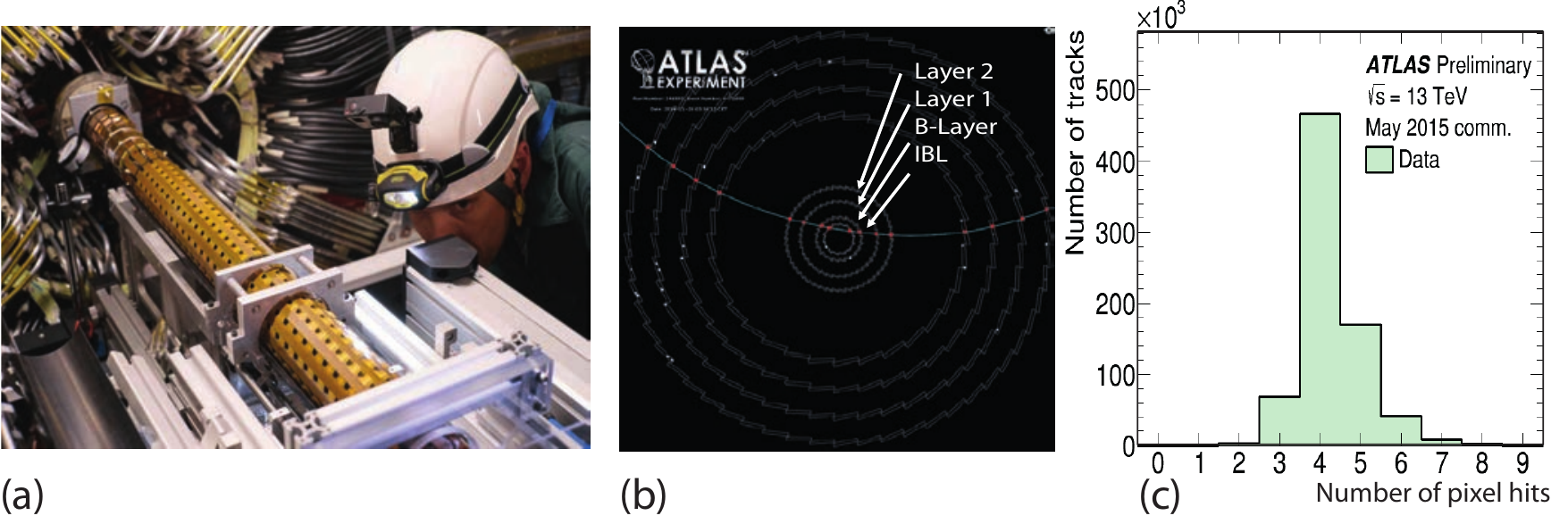}
\caption{Insertable B-Layer of ATLAS, a first pixel detector upgrade. (a) IBL insertion into ATLAS, (b) cosmic ray event and (c) 
hit distribution in 13 TeV data, both demonstrating the capability of this new four-hit pixel system.}
\label{fig:IBL} 
\end{figure}

The so-called Insertable-B-Layer (IBL) is the first upgraded pixel detector added in 2014 as an additional innermost pixel layer in
the ATLAS pixel system, now consisting of four layers (see fig.\,\ref{fig:IBL}). 
Compared to the existing 3-layer system of ATLAS, a large development step in terms of sensors and electronics has been undertaken in the context of the design and building of the IBL. Facing a severe hit efficiency degradation with the existing readout chip (FE-I3 chip \cite{FEI3}) already at LHC luminosities three times as high as the nominal one, IBL uses a new R/O-chip (FE-I4 \cite{GarciaSciveres:2011zz} in 130 nm technology), in which hits are stored locally in the pixels until triggered for readout, thus yielding a much higher efficiency.  Additional SEU hardening measures have also been implemented.
The pixel size is reduced to 50$\times$250\,$\mu$m. The FE-I4 chips can tolerate rates without too many hit losses up to several times the LHC design luminosity (10$^{34}$cm$^{-2}$s$^{-1}$), but it would yet be insufficient at rates as high as expected at the HL-LHC. IBL features two-chip pixel modules in size ($\sim$2$\times$4\,cm$^{2}$) almost as large as the 16-chip ATLAS pixel module using the FE-I3.
In addition to the R/O chip, IBL has 75\% planar silicon sensors with slimmed edges \cite{IBL-planar} and 25\% 3D-Si sensors \cite{3DSi-1,3DSi-2}, located at the ende of the IBL-staves, which are operating for the first time in a large pixel detector.

The demands of HL-LHC in terms of rate and radiation are currently addressed by the R\&D-Collaboration RD53 aiming for a 
pixel R/O chip in 65\,nm CMOS technology \cite{RD53}.  This chip can be used in hybrid pixel concepts and 
will be capable of digesting a hit rate of 1 GHz/cm$^2$ and  tolerate the ionising doses and fluences shown in table\,\ref{tab:accelerators}.

\subsection{Advantages and limitations of Hybrid Pixels}
Due to its split functionality (sensor and chip) the hybrid pixel technology has many advantages, especially in LHC-type
environments, due to the fact that both parts can be tailored to the demands, i.e. the sensor especially to optimally stand
irradiation and the readout chip to digest and process high rates ($\sim$MHz/mm$^2$). Complex signal processing, even already in the pixel cell, is possible, including for example, zero suppression and temporary storage of hits during the trigger L1 latency. Hybrid pixels are proven radiation hard to $10^{15}\, $n$_{eq}$/cm$^2$ and beyond. Spatial resolutions of the order of 10\,$\mu$m have been achieved.

However, the hybrid choice also has some severe disadvantages. They constitute a relatively large material budget, typically more than 3\% X$_0$ per layer 
in ATLAS and CMS. This is caused by the different components, sensor, R/O-chip, flex kapton, 
passive components, as well as support and cooling structures to operate at temperatures of  (-10\,\degree C), as well as services. The module production 
including bump-bonding and flip-chipping is complex and laborious, leading to a large number of production steps.
Consequently, hybrid pixel detectors are comparatively expensive.

Therefore, alternative pixel technologies have come into focus that aim to address the disadvantages mentioned above, while keeping the 
advantages to a large extent.  This development runs under the label  \emph{Depleted CMOS Pixels} or {DMAPS} (Depleted Monolithic Active Pixel Sensors). 

\section{DMAPS -- Monolithic CMOS Pixels for the HL-LHC Upgrade}
\subsection{The goal}
The goal of this new development is to employ commercial CMOS technology with some modifications to obtain sufficient signal and fast timing 
in monolithic CMOS designs, that merge sensor and R/O chip into one entity, while maintaining charge collection via charges drifting in an 
electric field inside the chip's substrate. In addition, the technology must survive the radiation environment at the HL-LHC, at least in the outer
layers, far enough from the interaction point such that the radiation levels are similar to those presently encountered at the inner layers, 
i.e.~100 Mrad dose and $2 \times10^{15} n_{eq}/$cm$^2$ particle fluence, respectively.

Simulations \cite{Hemperek_sim} have shown that these goals can be addressed by a) a high resistive ($>$1\,k$\Omega$cm) substrate material 
in which the charge is collected, b) a sufficiently high bias voltage (typically 20--50\,V) to supply the substrate with a high electric field and a reasonably large depletion volume according to 
$$
d \propto \sqrt{\rho \, V} 
$$ 
where $\rho$ is the resistivity and $V$ is the bias voltage, and by c) a large fill factor $F$ (the ratio of the charge collection area to the illuminated pixel area). Typical
numbers are $\rho$\,=\,2\,k$\Omega$cm, $V_{bias}$ = 20\,V and $F$ = 75\%.
In addition, one has to ensure that full CMOS circuitry (i.e. a balanced usage of PMOS and NMOS transistors in the electronics layer) is maintained by sufficient 
shielding of -  e.g. in an n-substrate - the PMOS transistors in n-wells shielded by deep p-wells by use of triple/quadruple well technology as shown in 
fig.~\ref{fig:HRCMOS_A&B}.  

Technologies that enable such designs are meanwhile available:
\begin{itemize}
\item {\bf HV add ons:} Special processing add-ons (from automotive and power management applications) that increase the voltage handling capability and create a depletion layer in a well's pn-junction of the order of 10-15\, $\mu$m. 
\item {\bf high resistivity wafers:} 8 inch high/mid resistivity silicon wafers accepted and qualified by the foundry. A depletion layer develops due the high resistivity.
\item{\bf special technology features:} Radiation hard processes with multiple nested wells are needed and a foundry must accept some process/DRC changes 
in order to optimise the design for HEP-applications.
\item {\bf backside processing:} Wafer thinning from the backside and a backside implant allow the fabrication of a backside contact after CMOS processing thus yielding
a strong electric field and uniform charge collection inside the sensor. 
\end{itemize}
The interest in CMOS pixels for the HL-LHC has risen due to the potential of low cost, large area devices that can cover the outer layers of ATLAS or CMS.
For the inner layers the possibility to achieve smaller pixels by in-pixel decoding  \cite{Peric:2015ska} is attractive. Hence the goal for CMOS pixels suited for LHC pp-collisions is designing an intelligent CMOS pixel cell with some depletion depth (e.g. 40--80\,$\mu$m) yielding a reasonably large signal ($\sim 4000 e^-$) with fast and in-time efficient  charge collection, while avoiding a long collection path on which charges can be trapped. 

\begin{figure}
\centering
\includegraphics[width=0.6\linewidth]{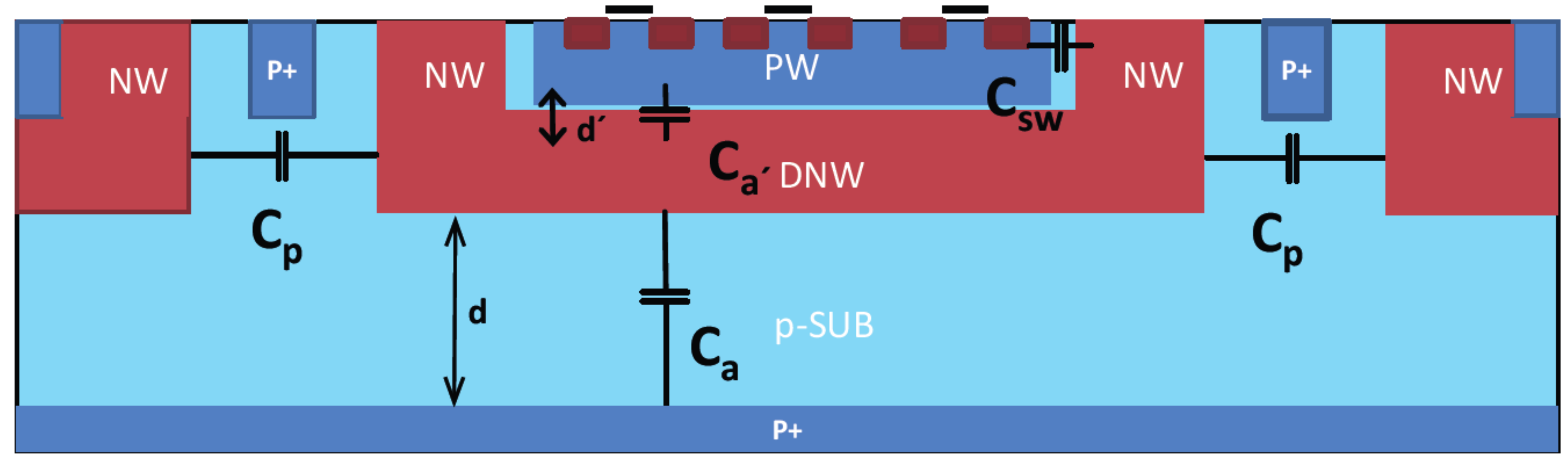}
\caption{Capacitance contributions of a typical CMOS pixel cell to the amplification node. Apart from capacitances to the backside and between pixels
a non-negligible contribution comes from the close by wells, deep n-well and p-well \cite{Krueger_priv}. }
\label{fig:capacitances} 
\end{figure}

Three current development lines can be identified: (a) HV-CMOS pixels (high voltage), (b) HR-CMOS pixels (high resistivity), and (c) CMOS on SOI pixels. They will be described in some more detail below. An important aspect to take note of regarding all CMOS pixel developments, is the fact that the capacitance of a 
CMOS pixel cell seen at the input of the preamplifier has contributions from various boundary structures (fig. \ref{fig:capacitances}). While for standard planar
pixel structures the total capacitance is dominated by the interpixel capacitance, for CMOS pixels a large contribution comes 
from the capacitance $C_a'$ between the collecting deep n-well and the p-well sitting above (see fig.\ref{fig:capacitances}) \cite{Krueger_priv}. 
Their close vicinity renders the capacitance contribution non negligible. Apart from increasing the total input capacitance, $C_a'$ also acts as the capacitance through which any activity on the bias lines can couple via the electronics into the sensor. 

\subsection{HV-CMOS pixels}
\begin{figure}
\centering
   \subfigure[HV-CMOS principle]{
    \includegraphics[width=0.47\textwidth]{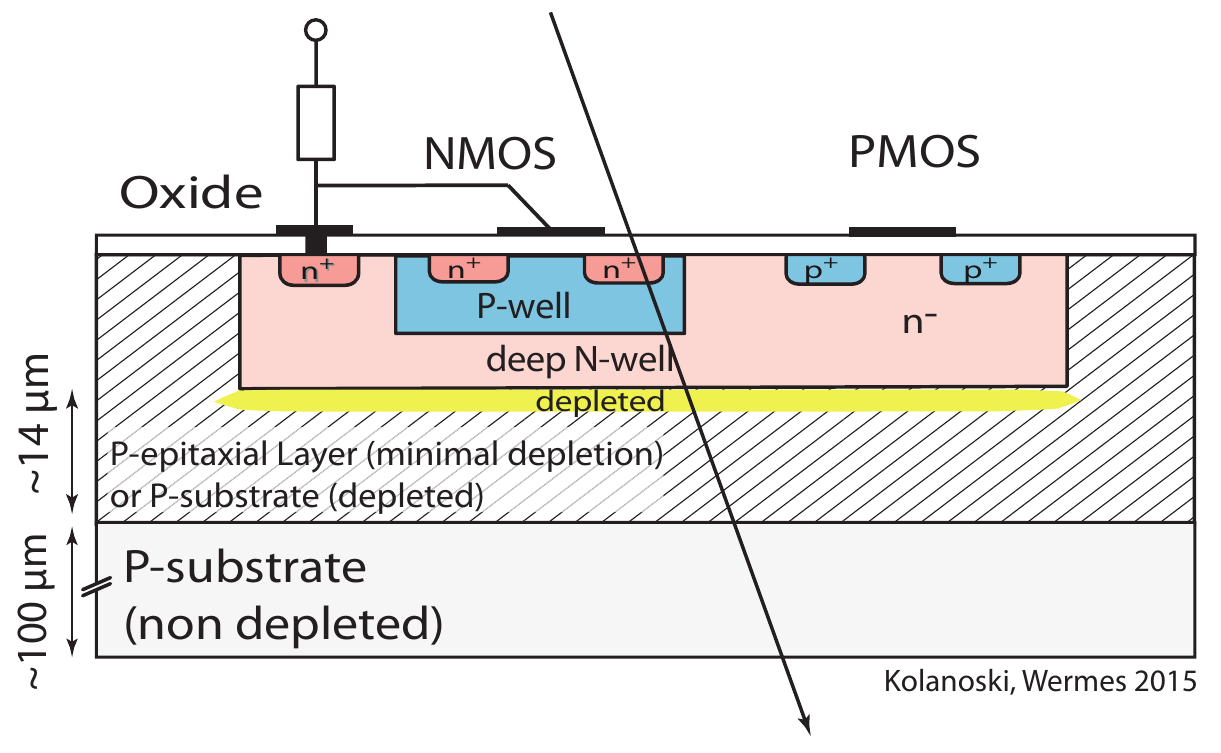}
    \label{fig:HVCMOS} }
     \subfigure[Efficiency]{
    \includegraphics[width=0.47\textwidth]{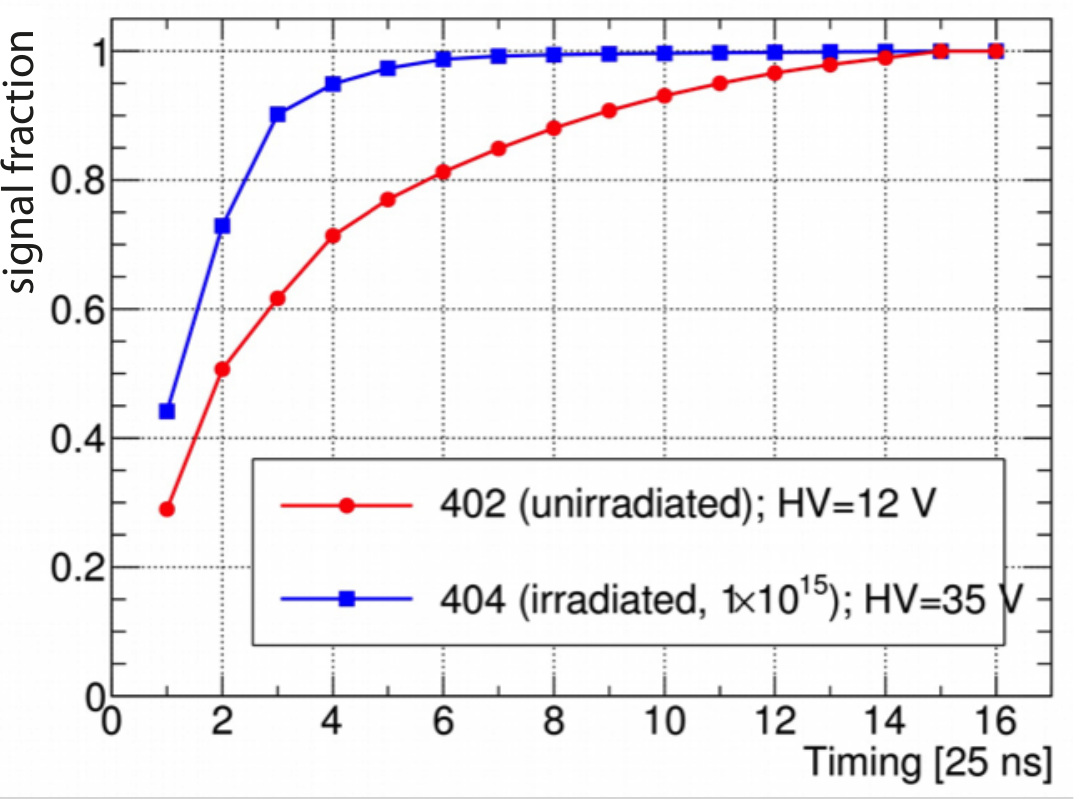}
    \label{fig:HVCMOS_results}}
\caption{(a) Sketch of a generic HV-CMOS pixel structure (from \cite{HKNW_book}). Both transistor types are embedded in a large deep n-well. The process has HV-add-ons which allow bias voltages up to $\sim$100 V. (b) Time delay of the signal when crossing a threshold is shown in units of 25 ns for an unirradiated and
an irradiated ($1\times 10^{15} n_{eq}/$cm$^2$) device. The HV-CMOS sensors was bonded with glue to a FE-I4 chip.}
\end{figure}

CMOS pixels with some depletion depth have first been implemented by Peric \cite{peric:2007} using a HV-technology with 350 nm and later 180 nm feature size. 
Both transistor flavours sit in a large deep n-well which at the same time acts as the charge collection node (fig.\,\ref{fig:HVCMOS}). 
The achievable depletion depth is expected around 10--20\,$\mu$m, but measurement on prototypes suggest even larger values. 
The output is an already amplified signal. One, currently much discussed, possibility is to directly (via bump bonds) or capacitively (via glue bonds) couple the 
pixel cells to the ATLAS front-end chip FE-I4  \cite{GarciaSciveres:2011zz}
as a second amplifying and fast processing device: an active hybrid pixel detector called CCPD (capacitively coupled pixel detector).   
HVCMOS detectors have survived TIDs of as much as 1\,Grad and fluences of 10$^{15}$ n$_{eq}$/cm$^2$. 
The signal after 1\,Grad is 1500 e$^-$ at a noise of about 60\,e$^-$. The time-integrated efficiency drops from 99\% (before irradiation) to 96\% after proton irradiation to 10$^{15}$ n$_{eq}$/cm$^2$. The signal rise time is typically 100\,ns, still too slow to achieve the required high intime efficiency as shown in 
fig.~\ref{fig:HVCMOS_results}. It is observed that the timing performance improves after irradiation (blue curve in fig. \ref{fig:HVCMOS_results}), most likely due to acceptor removal such that the fraction of charge collected by drift rather than diffusion becomes larger.

By connecting three CMOS subpixels to one FE-I4-pixel one can achieve an effective pixel size of 33$\times$125$\,\mu$m via pulse shape decoding \cite{Peric:2015ska}.   

\subsection{HR-CMOS pixels}
\begin{figure}
\centering
   \subfigure[HR-CMOS principle: variant-A]{
	\includegraphics[width=0.43\linewidth]{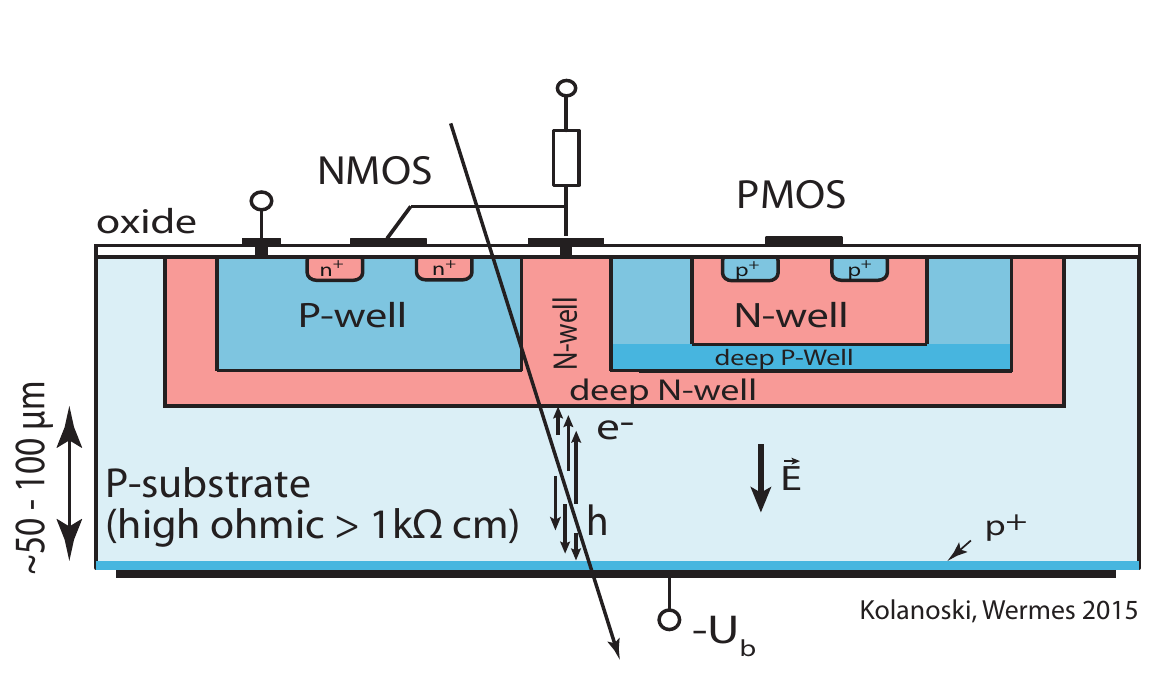}
    	\label{fig:HRCMOSA} }
   \subfigure[HR-CMOS priciple: variant-B]{
	\includegraphics[width=0.50\linewidth]{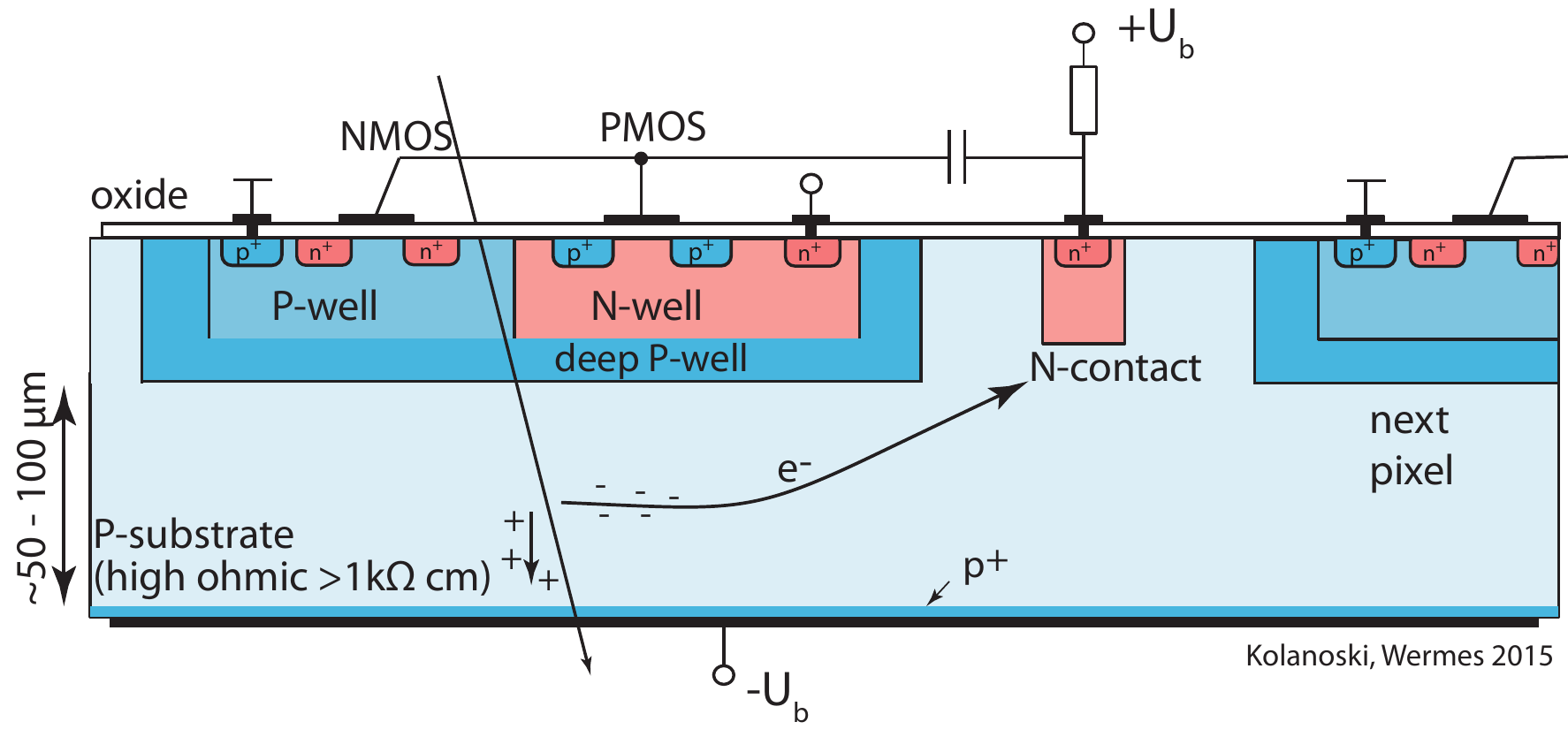}
    \label{fig:HRCMOSB}}
\caption{Structure of a HR-CMOS pixel cell. (a) Variant with a large collection n-well and short drift path; (b) Variant with a small collection n-well and a long drift path 
(from \cite{HKNW_book}).}
\label{fig:HRCMOS_A&B}
\end{figure}

Another approach uses high-resistive CMOS technologies with up to four well structures is shown in fig.~\ref{fig:HRCMOS_A&B}. In a $>$1k$\Omega$ 
resistive substrate material several nested wells provide charge collection as well as individually shielded transistors. In fig.~\ref{fig:HRCMOSA}
the charge collecting n-well is placed outside the electronics area providing a small capacitance at the expense of a long drift path. In fig.~\ref{fig:HRCMOSB}
the collecting deep n-well also houses the electronics similar to the HV-CMOS approach, however an additional deep p-well is employed isolating the 
n-well which houses the PMOS transistors from the collecting n-well. It is obvious that variant-B will benefit from the small capacitance of the collecting node
while it will likely be less tolerant against bulk damage than variant-A due to the different path length.  

\begin{figure}
\centering
   \subfigure[HR-CMOS pixel cell]{
		\includegraphics[width=0.48\linewidth]{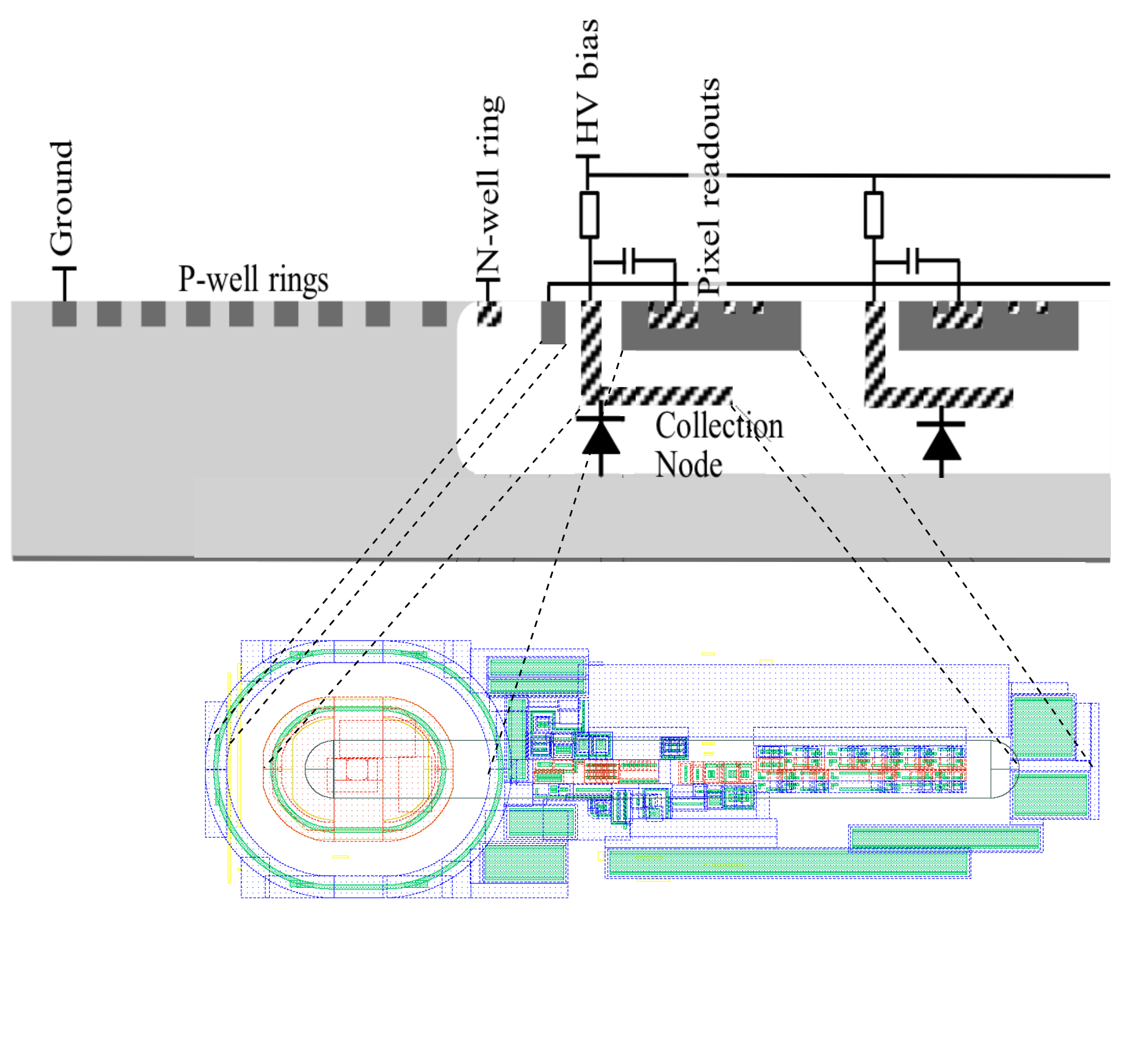}
   	\label{fig:LF_version-B} }
   \subfigure[HR-CMOS priciple: variant-b]{
	\includegraphics[width=0.45\linewidth]{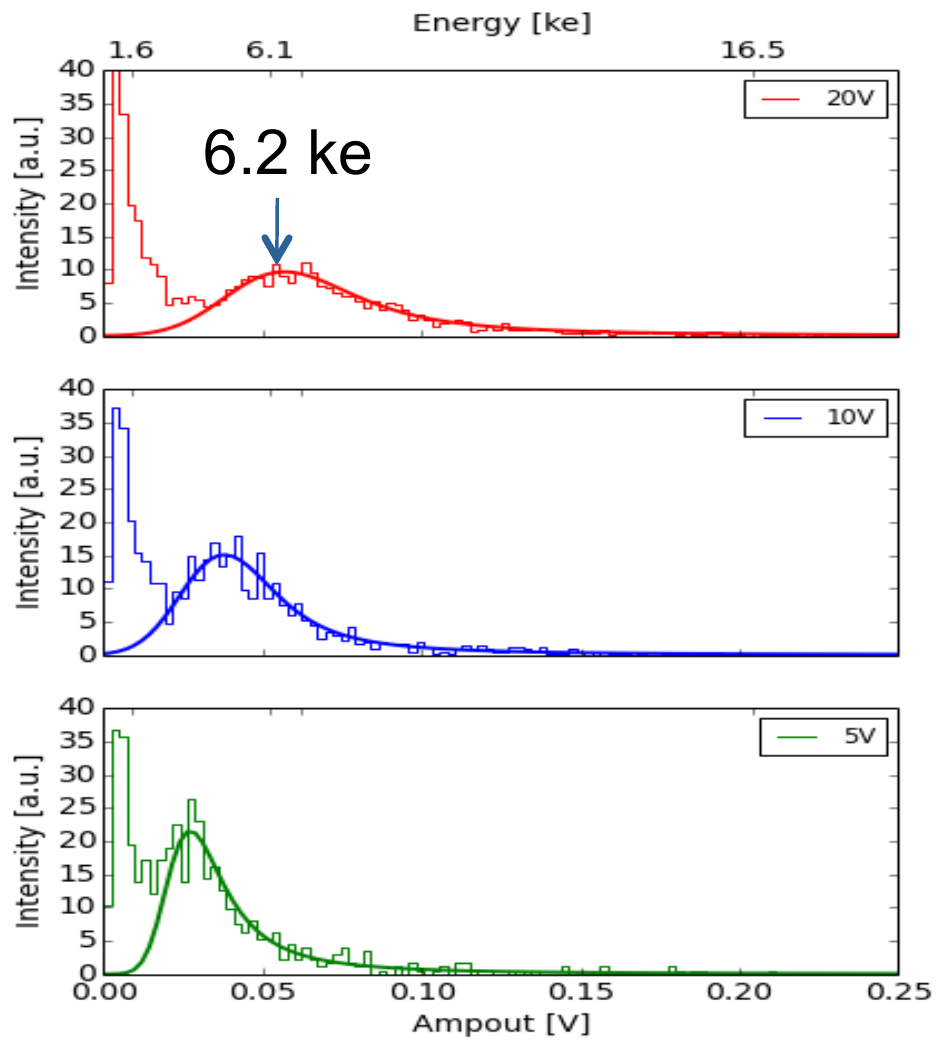}
    \label{fig:LF_ELSA}}
    \caption{Variant of an HR-CMOS pixel cell with reduced capacitance. (a) cell layout in two cross sections, (b) spectrum from 3.2 GeV electrons 
    at bias voltages of 5\,V, 10\,V, and 20\,V. The corresponding scale of energy deposited in the detector is shown on the top.}
\label{fig:HRCMOS_LFtests}
\end{figure}

HR-CMOS pixels in both variants of fig.~\ref{fig:HRCMOS_A&B} have been realised \cite{Havranek:2014ora,Obermann:2015oda} in different technologies offering high resistive substrates, multiple wells and in some cases also a backside contact. The latter is important, although not mandatory, to provide an electric field inside the sensor which is directed from the back to the collecting nodes. Figure \ref{fig:LF_version-B} shows the layout of a cell produced in 150 nm technology (LFoundry) on 2\,k$\Omega$\,cm substrate with a deep n-well structure aiming for smaller capacitance and faster rise time. A version of the cell with an n-well geometry encompassing the entire structure has also been realised. 

In fig.~\ref{fig:LF_version-B} the signal is shown after exposing the LF pixel array to a 3.2 GeV electron beam at the ELSA accelerator \cite{Hirono_iWORiD}. 
At a bias voltage of 20\,V a mean charge of 6200 e$^-$ is observed corresponding to about 60\,$\mu$m depletion depth and confirming that the bulk resistivity is about 2\,k$\Omega$\,cm. In-beam timing measurements show that the fraction of in-time hits is 91\% with a high threshold of 2600\,e$^-$, a very low threshold of 190\,e$^-$ yields a fraction of 79\% \cite{Hirono_iWORiD}.
In another process (ESPROS, 150 nm) HR-structures including thinning and a backside implant have been prototyped (see \cite{Havranek:2014ora,Obermann:2015oda}).

\subsection{CMOS-on-SOI pixels}
The Silicon-on-Insulator technology provides a buried oxide layer (BOX) which separates the 
CMOS electronics from the substrate layer. Both parts are connected by vertical via structures reaching through the BOX  and 
leading to an n-implant which acts as the charge collecting node. Monolithic SOI-based pixel structures have been 
developed for some time using the so-called fully depleted SOI technology \cite{Arai:2011ara}, invented for high speed 
CMOS electronics with reduced (parasitic) capacitances.  The CMOS electronics layer is embedded 
in depleted silicon. The pixel developments so far \cite{Arai:2011ara}  suffered from effects inherent to the BOX oxide layer, most notably 
the so-called backgate effect and from radiation effects. There have been successful measures to overcome these drawbacks \cite{Miyoshi:2013nwa} for moderate
radiation doses; FD-SOI pixels are nevertheless not suited for the conditions expected at the HL-LHC. 

\begin{figure}[h]
\centering
   \subfigure[SOI-MAPS-Structure]{
    	\includegraphics[width=0.34\linewidth]{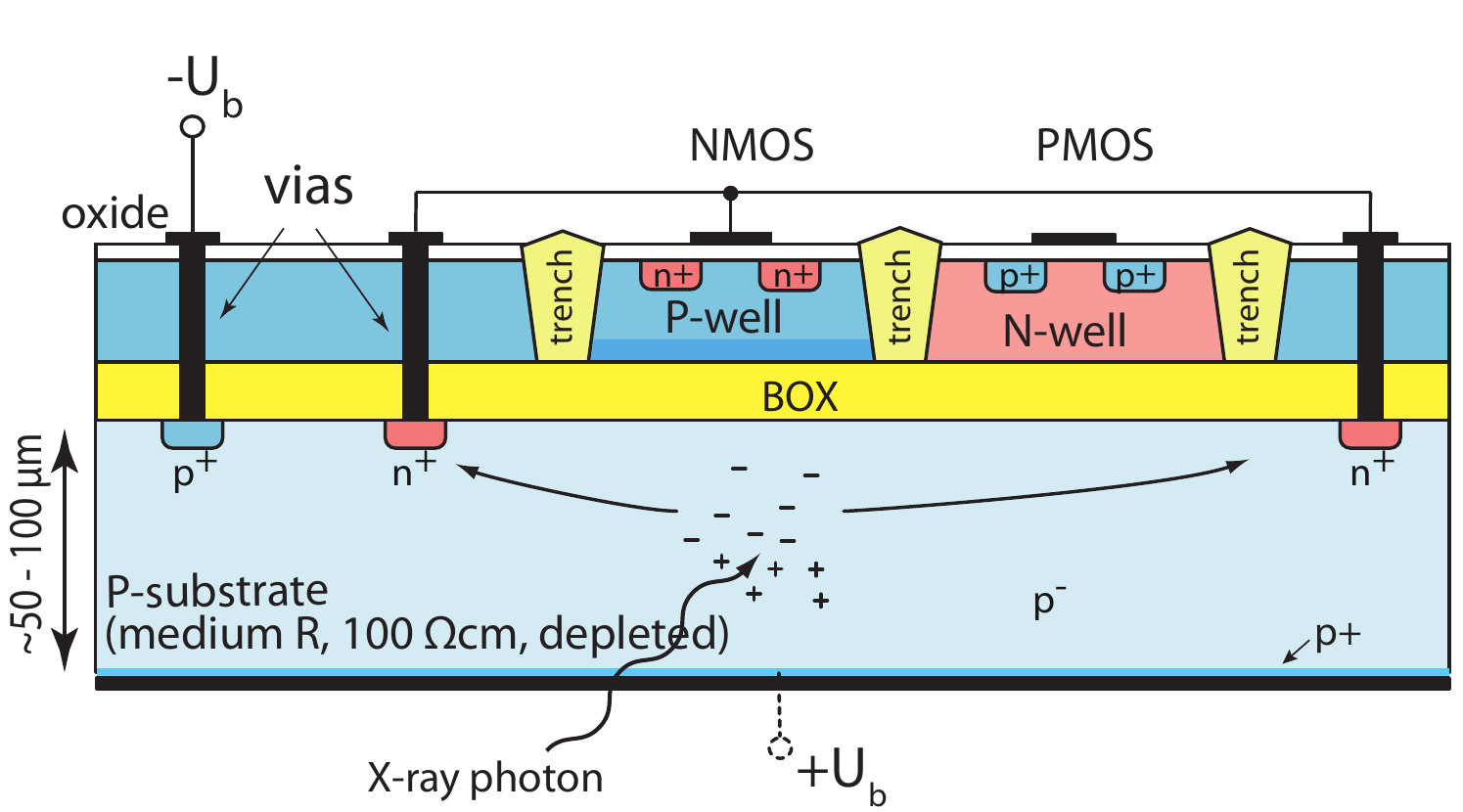}
  \label{fig:DMAPS_SOI} }
   \subfigure[Spectra]{
	\includegraphics[width=0.28\linewidth]{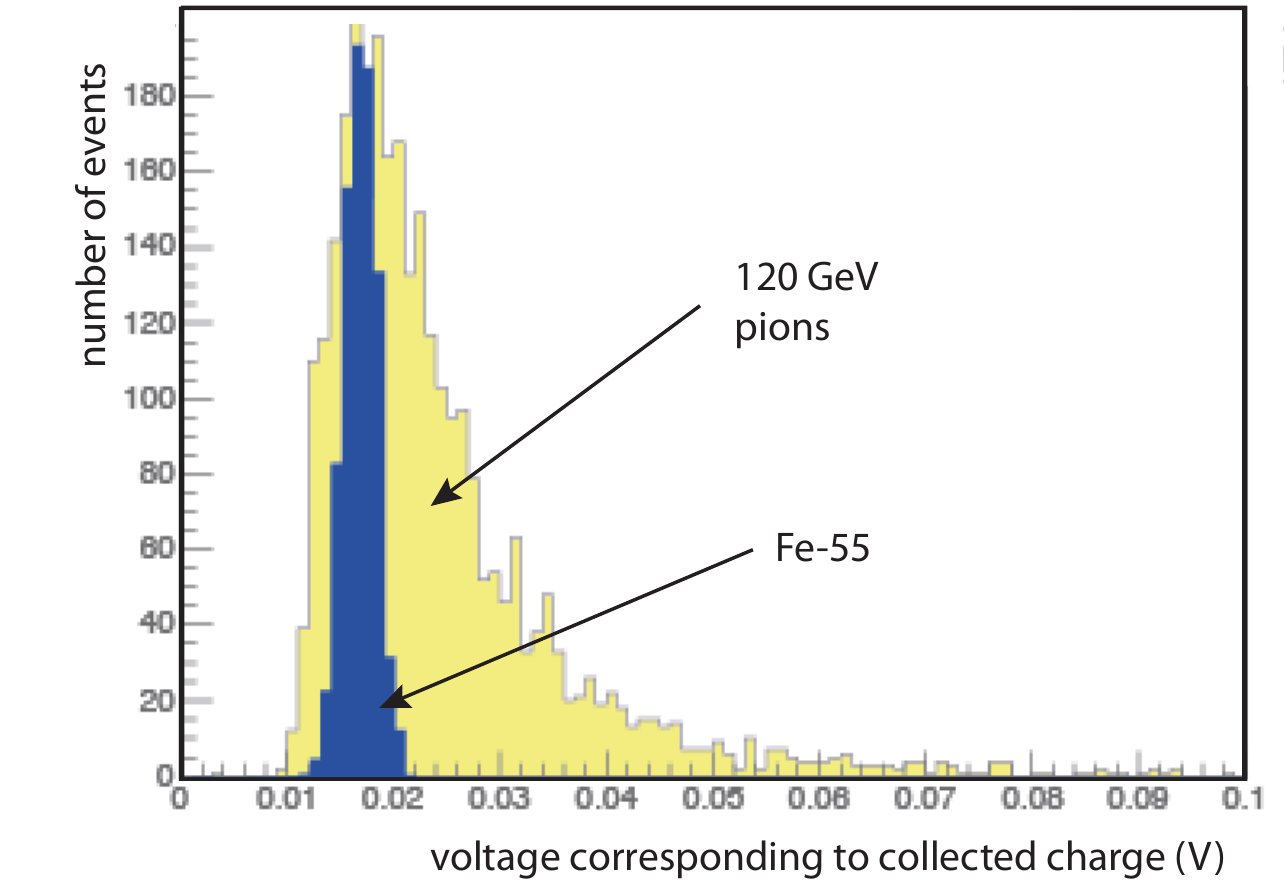}
    \label{fig:SOI_spectra}}
\subfigure[Irradiation NMOS]{
	\includegraphics[width=0.32\linewidth]{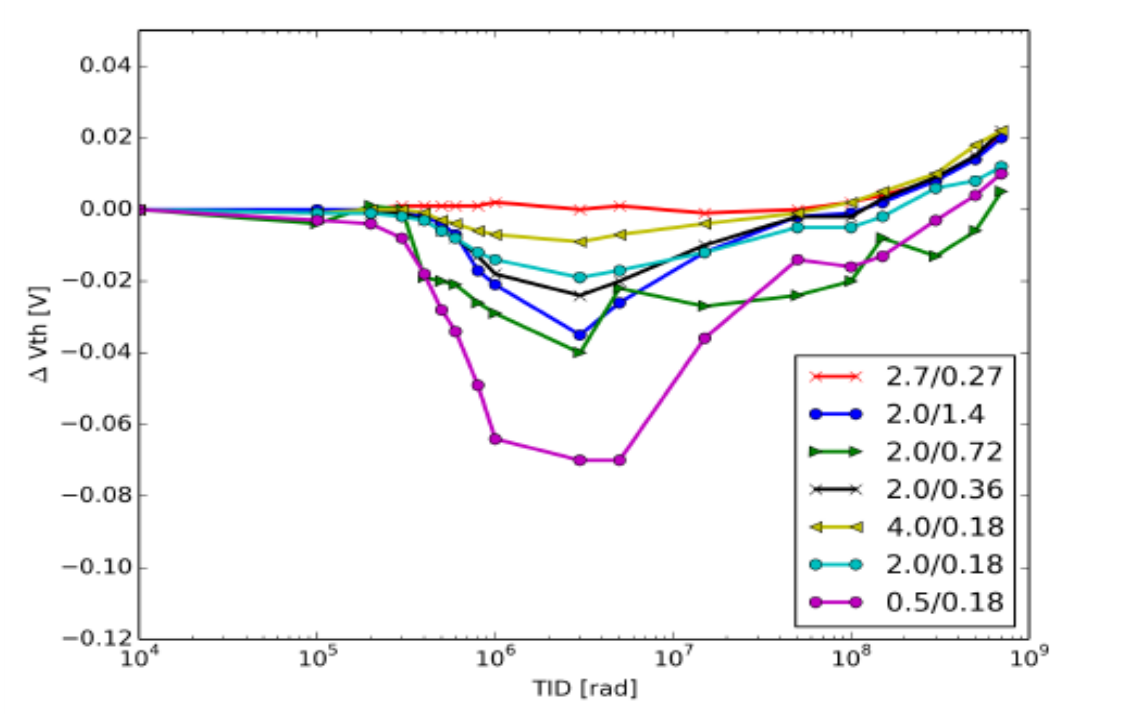}
    \label{fig:SOI_NMOS}}
\caption{Depleted SOI MAPS: (a) Pixel structure on thick film partially depleted SOI. The depleted charge collection volume is separated from the electronic layer by the BOX. The transitors are isolated by tranches and shielded by doped silicon at the top part of the BOX. (b) Signal spectrum of (unirradiated) SOI pixel structures from $\,^{55}$Fe (6\,keV) and 120 GeV pions. (c) Threshold shifts of the NMOS transistor in SOI technology as a function of X-ray dose.}
\label{fig:DMAPS-SOI}
\end{figure}

Thick film SOI, featuring trench isolation and doped, only partially depleted regions underneath CMOS transistors and thus shielding the transistors
(cf. fig\ref{fig:DMAPS-SOI}) are free of these difficulties and can also sustain mach larger radiation doses. 
SOI CMOS pixel detectors have been realised by \cite{Hemperek:2014yoa} and characterised  \cite{Sonia-FP2015} in the lab and in test beams. 
Figure\ref{fig:SOI_spectra} shows the signals obtained with an SOI pixel matrix (XFAB, 180nm) from $^{55}$Fe 
as well as from a beam of 120 GeV pions. The signal peaks at about 1700 e$^-$ corresponding to a depletion depth of 31$\,\mu$m.

Figure~\ref{fig:SOI_NMOS} shows threshold shifts of the NMOS transistor of these SOI structures with X-ray radiation \cite{Sonia-FP2014}.
The maximum shift is 70 mV for NMOS transitors improving again towards higher doses because the damage at the Si-SiO$_2$ interface compensates
the effect of the oxide damage. PMOS transistors are not compensated; the shift after 1 Grad is about 100 mV. Both values
are tolerable for CMOS circuitry when such threshold shifts are taken into account in the design.   

\section{Conclusions}
Hybrid pixels have convincingly demonstrated their capability to cope with the high rate and high radiation environments at the LHC and even at the
planned HL-LHC. A first upgrade of such devices is the ATLAS Insertable B-layer (IBL). However, hybrid pixels also have a number of drawbacks, among them 
is the material and the production costs.
Intensive research and development of monolithic pixel detectors aiming to operate LHC-like environments 
is currently  considered an attractive alternative direction. R\&D profits from the rapid progress in micro electronics, meanwhile allowing a number of technology variants.
CMOS pixels could offer is a low cost potential which is especially attractive for large areas needed at the outer layers of the LHC experiments, where also
the radiation levels are less severe.
For the inner layers, closer to the interaction point, the possibility to obtain smaller pixel sizes than with hybrid pixels is attractive. 
\section*{Acknowledgments}
The author would like to thank his colleagues from the ATLAS CMOS Pixel collaboration. 
This work was supported by the Deutsche Forschungsgemeinschaft DFG, 
grant number WE 976/4-1
and by the German Ministry of Education and Research BMBF under 
grant numbers 05H2012PD1 and 05H15PDCAA.  


\end{document}